\documentclass[12pt]{article}
\begin{document}
\begin{titlepage}
\vtop{\hbox{hep-th/0011057}
\hbox{HU-EP-00/51}}
\vskip 3cm
\centerline{\Large\bf{T-duality and  Actions}} 
\centerline{\Large\bf{for Non-Commutative D-Branes}}
\vskip 1cm
\centerline{Radu Tatar\footnote{tatar@qft.physik.hu-berlin.de}}
\bigskip
\centerline{\it Humboldt-Universit\"at zu Berlin, Institut f\"ur  
Physik,}
\centerline{\it Invalidenstrasse 110, 10115 Berlin, Germany}
\smallskip
\vskip 4cm
\centerline{\bf Abstract}
We show how the T-duality is realized for D-branes with 
noncommutative world-volume coordinates. We discuss D-branes wrapped on 
tori and the result is that
the recently found noncommutative actions form a 
consistent collection due to the T-duality mapping between 
noncommutative D-branes and rotated commutative D-branes on deformed tori. 
\bigskip
\end{titlepage}
\section{Introduction}
Recently, the actions for the BPS and non-BPS D-branes have been discussed
extensively in order to study their dynamics. By turning on constant
NS-NS 2-form $B$-field along the world-volume of either type of  D-branes, 
the world-volume
action becomes a noncommutative field theory \cite{sch,sw,gms,dmr,kra2,gms1,
sei,kra3,sen1,gms2,kra4}. 
The product of the fields of the non-commutative field theories is a
$*$ product which involves a $\theta$ parameter related to the value of the
$B$ field.
Besides the $B$ field, 
we can turn on different magnetic fluxes $F$ on the D-branes and then the
non-commutative field theories on the D-branes have multiple 
$\theta$ parameters related to the different values of $F + B$ 
\cite{ot5,nappi,dasy}.

In \cite{ms}, the Chern-Simons couplings for D-branes were generalized to 
the case of branes with noncommutative world-volume coordinate in a manifestly
background independent way. For D-branes with $B$ and $F$ fields, this 
includes a factor of $\frac{B}{F+B}$ in front of the usual coupling between 
the $B$ and $F$ fields and the RR-forms in type II theories. 

One important check of the results of \cite{ms} is to study the transformation
of the noncommutative Chern-Simons under T-duality.  
This is the subject of the present paper. We consider non-commutative 
D-branes wrapped on tori with different dimensions and we check the 
transformation of their actions under
T-dualities with respect to different directions on the tori. 
We use the results of \cite{chu,chen,ima,blu1,blu2,blu3} 
concerning the effect of
T-duality on the noncommutative D-branes. 
The results is that, after T-duality, a 
noncommutative D-brane becomes a commutative D-brane rotated with respect to
the original coordinates of the torus and the torus itself becomes deformed.
The rotation of the brane is related to the value of $B + F$ and the 
deformation of the torus is related to the value of $B$.  

By starting with a non-commutative D9-brane, we map the Chern-Simons terms
obtained 
via double and direct dimensional 
reduction as done previously for commutative 
BPS and non-BPS D-branes in \cite{al,b1,ga,b2,jm}.

\section{T-duality and Chern-Simons terms For Noncommutative D-branes}
We check the consistency of the Chern-Simons terms for
non-commutative D-branes discussed on \cite{ms}. 
We start with a D9 brane with two compact directions 
on a $T^2$ torus, with NS 2-form $B$  and magnetic flux $F$ turned on the
directions of the torus.
Therefore the field theory on the brane becomes non-commutative. 

As described in \cite{ms}, one possible choice for the values of the fields is
to take $F + B = Q^{-1}$ constant and to allow a variable value for 
$B = \theta^{-1}$. In this case the Born-Infeld and Chern-Simons actions could
be written as:
\begin{equation}
\label{dbi1}
 S_{DBI} =  T_9  \int d^{p+1} x~ {\mbox{Pf}~Q\over \mbox{Pf}~\theta} 
~\sqrt{\det\big(g_{ij} + (F + B)_{ij}\big)}
\end{equation}
and 
\begin{equation}
\label{cs1}
 S_{CS} = \mu_9 \int_x {\mbox{Pf}~Q\over \mbox{Pf}~\theta}
~\sum_n C^{(n)} e^{F + B} 
\end{equation}
where $T_9$ and $\mu_9$ are the tension and charge of the D9 brane. In the
case of only 2 directions of B and F fields, we can write 
$\mbox{Pf}~Q = Q$ and $\mbox{Pf}~\theta = \theta$ 
so the above formulas become:
\begin{equation}
\label{dbi2}
S_{DBI} =  T_9  \int d^{2} x~ {Q \over \theta} 
 ~\sqrt{\det\big(g_{ij} + (F + B)_{ij}\big)}
\end{equation}
and 
\begin{equation}
\label{cs2}
S_{CS} = \mu_9 \int_x {Q \over \theta}
 ~\sum_n C^{(n)} e^{F + B} 
\end{equation}

Let us consider that the directions of the two-torus $T^2$ and of the $B$ and 
$F$ fields are on the $(x^1, x^2)$ plane. Therefore, if we want to
discuss the action of T-duality, we need to distinguish between T-duality on
the compact $(x^1, x^2)$ directions and T-dualities with respect to any other 
possible compact direction. 

If we consider $x^3$ as a compact direction and take a T-duality with respect
to it, then the D9 brane will become a D8 brane wrapped on  $(x^1, x^2, x^3)$,
and the action will be:
\begin{equation}
\label{dbi3}
S_{DBI} =  T_8 \int d^{2} x~ {Q \over \theta} 
\sqrt{\det\big(\hat{g}_{ij} + (\hat{F} + \hat{B})_{ij}\big)}
\end{equation}
and 
\begin{equation}
\label{cs3}
S_{CS} = \mu_8 \int_x {Q \over \theta}
\sum_n \hat{C}^{(n)} e^{\hat{F} + \hat{B}}
\end{equation}
where $T_8$ and $\mu_8$ are the tension and charge of the D8 brane.
In this case, the 9-th component of the BI vector $A_{3}$ gets mapped 
into the 
transverse dimension $\phi_{3}$ to the D8 brane in the usual sense.

What happens now if we consider a T-duality with respect to $x^1$ or $x^2$
directions? As discussed in 
\cite{chen,ima,blu1,blu2,blu3}, there are two things
which happen: the $T^2$ torus gets deformed and the D8 branes is rotated in
the  $(x^1, x^2)$ plane. If we consider a T-duality along the 
$x^2$ direction, we obtain a torus with a rotation of the $x^2$ axis
into an $x'^2$ axis by an angle $\pi/2 - \alpha$ given by:
\begin{equation}
\label{angles1}
\mbox{cot}~\alpha = B = \theta^{-1}
\end{equation}
(so it makes and angle $\alpha$ with the $x^1$ axis) 
and the D8 brane has a direction on the deformed torus $(x^1, x'^2)$ 
at an angle $\beta$ with respect to the direction $x^1$ where
$\beta$ is given by:
\begin{equation}
\label{angles2}
\mbox{cot}~\beta = B + F = Q^{-1}
\end{equation} 

We want to see the consistency of T-duality at the level 
of Chern-Simons action.
From the formula (\ref{cs3}), we see that for the D9 brane
 we have the couplings
$C^{(10)} + C^{(8)} \wedge (F + B)$ when the $F, B$ are 
in the $(x^1, x^2)$ directions and
$\hat C^{(8)}$ is in the $(x^0,x^3\cdots,x^9)$ directions. 
Together with the factor $\frac{Q}{\theta}$ this will give
\begin{equation}
\label{d9}
 \int \frac{Q}{\theta}~(C^{(10)} + C^{(8)} \wedge (F + B)) 
\end{equation}

A double dimensional reduction of the D9 brane on the $x^2$ direction gives a 
commutative D8 brane in the $x^1,\cdots,x^{10}$ directions with 
Chern-Simons terms containing the term
\begin{equation}
\label{dr}
\int  \frac{\mbox{tan}~\alpha}{\mbox{tan}~\beta}~(C^{9} + C^{8} d \chi)  
\end{equation}
where the nine-dimensional scalar field comes from
the component of the BI vector in the direction $x^2$ over which we reduce.

We then consider a direct dimensional reduction of a commutative D8 brane 
on the  $(x'^{2},\cdots,x^{10})$ directions whose Chern-Simons term in 
10 dimensions does not contain the   $Q \over \theta$ term in the action.
After reduction, it has a Chern-Simons terms 
containing the sum 
\begin{equation}
\label{dir}
 \int  (C^{9} + C^{8} d \chi') 
\end{equation}
where $\chi' = \Phi^{2}$ comes from the reduction on the transverse direction.

The integral in (\ref{dr}) is taken over $d x^1$ and in (\ref{dir}) is taken
over $d x'^2$.  But $\frac{d x'^2}{d x^1}$ is just $\mbox{tan}~\alpha$ so we
see that we see the appearance of $\mbox{tan}~\alpha$ in (\ref{dir}).
Moreover, the directly dimensional reduced D8 brane is at an angle 
$\beta$ with respect to $x^1$ so there is a factor $\mbox{cot}~\beta$ also
appearing in the action. Therefore we have a mapping between 
the 2-nd component of the BI vector and the extra transverse 
direction.

What happens if we take two T-dualities in the $x^1, x^2$ directions?
The D9 brane becomes a commutative D7 branes. 
In \cite{chu,blu1} is has been argued that the coordinates
$x^1, x^2$ of the D7 brane do not commute in this case. 
By turning on $B_{12}$ field and keeping the other 8 directions non-compact,
this induces D7 - branes on the world-volume of the D9-branes, the D7-branes
must couple to the RR 10-form potential as discussed by Myers \cite{mye1}
and this induces a term like
\begin{equation}
[\phi^{1}, \phi^{2}] \wedge C_{12 i_1 \cdots i_8}
\end{equation}
The D7 branes obtained by T-duality were discussed to be generated by applying 
an asymmetric rotation to an ordinary D7-brane with pure Neumann or Dirichlet
boundary conditions \cite{chu,blu1}. 
The commutator of the transverse coordinates becomes in
the large $F, B$ limit as $[x_1, x_2] = 1/(F+B) = Q$ so if we identify
$\phi^{1} = x^1, \phi^{2} = x^2$, the Myers coupling becomes
 $Q_{12}~C_{12 i_1 \cdots i_8}$. 
There is also a term $1/\theta$ coming
into the action because of the deformation of the torus.
So the  the Chern-Simons term for the D7 branes will
contain a part of the form 
\begin{equation}
\label{d7}
\int \frac{Q}{\theta} \quad C_{12 i_1 \cdots i_8}.
\end{equation}
and terms with derivatives of the transverse directions.
If we now consider the term $\frac{Q}{\theta}~C^{(10)}$ in the D9 brane 
action, this will be mapped into the Chern-Simons term for the D7 brane
after the two T-dualities.

Let us now discuss the case when we have four compact directions which are
seen as $T^2 \times T^2$ on the $x^1, x^2$ and $x^3, x^4$ directions 
respectively. We turn on $B$ and $F$ fields such that 
$B_{12} = B_{1},~F_{12} = F_{1}$ and 
$B_{34} = B_{2},~F_{34} = F_{2}$.
Then we define 
\begin{equation}
Q_{i}^{-1} = B_{i} + F_{i}, i = 1, 2
\end{equation} 
and 
\begin{equation}
\theta_{i}^{-1} = B_{i}, i = 1, 2
\end{equation}
and we insert in formula (\ref{cs1}) $\mbox{Pf}~Q = Q_1~Q_2$
and $\mbox{Pf}~\theta = \theta_1~\theta_2$. 

We then consider two T-dualities
in the $x_2$ and $x_4$ directions. Under T-duality, the D9 brane goes 
into a D7 brane rotated in the $x^1, x^2$ plane at an angle
$\beta_{1}$ with respect to the $x^1$ direction and rotated in the 
$x^3, x^4$ plane at an angle $\beta_2$ with respect to the
$x^3$ direction. The angles are given by
\begin{equation}
\mbox{cot}~\beta_i = B_i + F_i = Q_i^{-1}, i = 1, 2.
\end{equation}
In the same time the direction $x^2$ is rotated by an angle
$\pi/2 - \alpha_1$ where $\mbox{cot}~\alpha_1 = Q_1$ 
and the direction $x^4$ gets rotated by an angle $\pi/2 - \alpha_2$
where $\mbox{cot}~\alpha_2 = Q_2$. 

For the D9 brane we have the coupling 
\begin{equation}
\sum_{i=1}^{2}  C^{(10)} + C^{(8)} \wedge (F_i + B_i)
\end{equation}  
so the Chern-Simons term is:
\begin{equation}
\int \prod_{i=1}^{2}~\frac{Q_i}{\theta_i}
~(C^{(10)} + \sum_{i=1}^{2}~C^{(8)} \wedge (F_i + B_i))
\end{equation}
or
\begin{equation} 
\int (\prod_{i=1}^{2}~\frac{Q_i}{\theta_i}~C^{(10)}
+ \sum_{i=1}^{2}~\frac{Q_i}{\theta_1~\theta_2}~C^{(8)}) 
\end{equation}
which is
\begin{equation} 
\int (\prod_{i=1}^{2}~\frac{Q_i}{\theta_i}~C^{(10)}
+ \sum_{i=1}^{2}~\mbox{tan}~\alpha_1~\mbox{tan}~\alpha_2
~Q_i~C^{(8)})
\end{equation}
A double reduction of the D9 brane in the $x^2$ direction gives a 
noncommutative D8 brane with a Chern-Simons coupling:
\begin{equation}
\label{dr1}
\int (\prod_{i=1}^{2}~\frac{Q_i}{\theta_i}~(C^{(9)} +
C^{(8)}~d \chi)
+ \mbox{tan}~\alpha_1~\mbox{tan}~\alpha_2
~Q_1~(C^{(7)} + C^{(6)}~d \chi)))
\end{equation}
We consider also a direct dimensional reduction of a noncommutative D8 
brane on $(x'^2,\cdots,x^{10})$ whose noncommutative 
Chern-Simons action in 10 dimensions is:
\begin{equation}
\sum_{i=1}^{2} \frac{Q_2}{\theta_2}~(C^{(9)} + C^{(7)} \wedge (F_i + B_i))
\end{equation}  
After reduction the Chern-Simons action becomes:
\begin{equation}
\label{dir1}
\int Q_2~\mbox{tan}~\alpha_2~(C^{(9)} + C^{(8)}~d \chi' + (C^{(7)} 
+ C^{(6)}~d \chi') \wedge (F_i + B_i))
\end{equation}
We now compare equations (\ref{dr1}) and (\ref{dir1}). We see that the 
difference is a factor $\frac{\mbox{tan}~\alpha_1}{\mbox{tan}~\beta}$.
But this is exactly coming from the deformation of the torus and from the
rotation of the D-brane in the $(x^1, x^2)$ plane as explained above. 
So again there is a mapping between the 2-nd component of the BI vector and
the extra transverse direction.
The same discussion applies if we consider a T-duality on the $x^4$ direction.

If we take two T-dualities in the $x^2, x^4$ directions, we need to compared
a twice doubly dimensional reduced noncommutative D9 brane and a twice 
directly dimensional reduced D7 brane. We can start directly in 9 dimensions 
with a non-commutative D8 brane and make only one reduction. We can start with
both formulas (\ref{dr1}) or (\ref{dir1}). If we start with  (\ref{dr1}) and
make a double dimensional reduction on the $x^4$ direction, we obtain a 
commutative D7 brane with a Chern-Simons action:
\begin{equation}
\label{dr2}
\int (\prod_{i=1}^{2}
~\frac{Q_i}{\theta_i}~(C^{(8)} + C^{(7)}~d~\tau + C^{(7)}~d~\chi + 
C^{(6)}~d~\tau~d~\chi))
\end{equation}
where $\tau$ is the eight-dimensional scalar field which comes from the
component of the nine-dimensional BI vector in the direction $x^4$ over which
we reduce.
We then consider a direct dimensional reduction of a commutative 
D7 brane on the $(x'^{2}, x'^{4},\cdots,x^{10})$ directions from ten to eight 
dimensions, this has a Chern-Simons term:
\begin{equation}
\label{dir2}
\int (C^{(8)} + C^{(7)}~d~\tau' + C^{(7)}~d~\chi' + 
C^{(6)}~d~\tau'~d \chi'))
\end{equation}
where $\tau' = \Phi^{4}$ comes from the reduction on the transverse direction.
The supplementary factor $\prod_{i=1}^{2}~\frac{Q_i}{\theta_i}$ 
in (\ref{dr2}) comes from the rotation of the two tori $T^2$ and from 
the rotation of the directions of the D7 brane in the $(x^1,x^2)$ plane and
in the $(x^3, x^4)$ plane, therefore we see that the two components
$\chi, \tau$ of the BI vector are mapped into the extra transverse directions
$\chi', \tau'$, as required by T duality.

We would like now to discuss the non-abelian case. In \cite{jm,mye1}, the
action for the case of multi-branes on top of each other is given and it
involves the replacement of partial derivatives for the transverse fields by
covariant derivatives. By starting from the D9 branes, after reduction to 9
dimensions, there is no other scalar field except the 10-th component of the
gauge field so we still have partial derivatives and the above discussion is
the same, the only difference being that the Chern-Simons action 
(\ref{cs1}) now writes:
\begin{equation}
\label{cs6}
 S_{CS} = \mu_9 \int_x {\mbox{Pf}~Q\over \mbox{Pf}~\theta}
\mbox{Tr}~\sum_n C^{(n)}~e^{F + B} 
\end{equation}
where by Tr we mean the symmetric trace description.

If we consider the case of $N$ D9 branes on 
$T^2$ in the $x^1, x^2$ directions with magnetic fluxes 
$F_{i},~i = 1,\cdots,N$, a T-duality in the $x^1$ direction would 
give a deformed $T^2$ torus by an angle $\mbox{cot}~\alpha = B$ and 
$N$ D8 branes rotated by angles 
$\mbox{cot}~\beta_i = B + F_i,i = 1,\cdots,N$.
In this case, we need to consider the spectrum of the open strings between the
different D8 branes which could contain tachyons so the system might be 
unstable. As discussed in \cite{chen,blu1}, D-branes at angles could become
unstable and behave as D $- \bar{\mbox{D}}$ systems.
It would be very interesting to have a complete discussion of these
phenomena. 

We can continue the above discussion for the case of 
six compact directions which 
seen as a product of three tori $T^2 \times T^2 \times T^2$ 
in the $(x^1, x^2)$,
$(x^3, x^4)$ and $(x^5,x^6)$ directions respectively. 
We turn on $B$ and $F$ fields such that 
 $B_{12} = B_1,~B_{34} = B_2,~B_{56} = B_3$ and
 $F_{12} = F_1,~F_{34} = F_2,~F_{56} = F_3$ . We then have three values for
$Q_i$ and three values for $\theta_i$ and $\mbox{Pf}~Q = Q_1~Q_2~Q_3$ and
$\mbox{Pf}~\theta = \theta_1~\theta_2~\theta_3$. 

By applying three T-dualities with respect to $x^2, x^4, x^6$, the factor 
$\frac{\mbox{Pf}~Q}{\mbox{Pf}~\theta}$ appears naturally when comparing the
double and direct direction and the formula (\ref{cs1}) stands also
for this case.

Another check for the formulas of \cite{ms} is made when 
we start with a D7 brane on $(x^1,\cdots,x^8)$ directions with
$B$ and $F$ fields on the $(x^1, x^2)$ directions. As discussed in \cite{ms},
we have the transverse coordinates $\Phi^8, \Phi^9$ which are functions of the
noncommuting brane coordinates so they do not commute. 
The Chern-Simons term will now look like:
\begin{equation}
\label{csd}
S_{CS} = \mu_7 \int_x {Q \over \theta}~P[e^{i(i_{\Phi} * i_{\Phi})}
\sum_n C^{(n)}]~e^{Q^{-1}} 
\end{equation}
where P represents the pullback of the transverse brane coordinates and
$Q^{12} = (F^{12} + B^{12})^{-1}$. 
The exponential term $e^{i(i_{\Phi} * i_{\Phi})}$
just implies the appearance of a Myers-type term like 
$Q^{9,10} = - i~[\Phi^{9}, \Phi^{10}]$. 

If we use the above discussion, a T-duality on the $x^1$ direction would give
a rotated D6 brane in a tilted torus. The D6 brane is now commutative so the
transverse directions are functions of commuting coordinate and they 
commute. This would mean that in this case we do not have
Myers-type term because $[\Phi^{9}, \Phi^{10}] = 0$, and this is expected
because the Chern-Simons term for the D6 brane involves only angles of
rotation which are related to $Q^{12}$ and $\theta^{12}$ and not to 
$Q^{9,10}$.

The non-BPS D-branes could be treated in a similar fashion by 
using the results of 
\cite{ga,b2} concerning the T-duality 
for commutative non-BPS D-branes. A formula has been 
proposed in \cite{ms} for a non-BPS D8 brane as
\begin{equation}
S_{CS} = \frac{\mu_{8}}{2T_{min}}\int  {\mbox{Pf}~Q\over \mbox{Pf}~\theta}\,
{\cal D}T\, C^{(n)}~e^{Q^{-1}}
\end{equation}
where ${\cal D}_i T = -i~(Q^{-1})_{ij}~[X^j,T]$ is a covariant derivative 
which is background independent and linear in $[X^j,T]$.

Another important example of non-commutative field theories appear when the
D-branes are not wrapped on tori but on $S^2$ cycles.
When studying D3 branes orthogonal to 
orbifolds or conifolds, the D5 branes wrapped on the resolution vanishing
2-cycles have naturally 
$B$ and $F$ fields on their world-volume and this implies  
Chern-Simons couplings which induce fractional D3 brane charges
\cite{do1,do2,kar,gi,dm2,k1,ot3,bere,ot4}.   
In the case of non-BPS systems of branes, by turning on different $F$ fluxes on
different branes we can make the system stable as discussed in 
\cite{oz,ot5,kr1,li}. 

\vskip 3cm

\centerline{{\Large \bf Acknowledgements}}

\vskip 1cm

We would like to thank Eric Bergshoeff and Ralph Blumenhagen for discussions
and to Dieter Lust and Keshav Dasgupta for comments on the manuscript.
The work was supported by DFG.

\end{document}